\documentclass[aps,twocolumn,showpacs,floatfix,superscriptaddress]{revtex4}
\usepackage{graphicx}
\usepackage{psfrag}
\begin{document}
\newcommand{\beq}{\begin{equation}}
\newcommand{\eeq}{\end{equation}}
\newcommand{\ben}{\begin{eqnarray}}
\newcommand{\een}{\end{eqnarray}}
\newcommand{\bea}{\begin{array}}
\newcommand{\eea}{\end{array}}
\newcommand{\om}{(\omega )}
\newcommand{\bef}{\begin{figure}}
\newcommand{\eef}{\end{figure}}
\newcommand{\leg}[1]{\caption{\protect\rm{\protect\footnotesize{#1}}}}
\newcommand{\ew}[1]{\langle{#1}\rangle}
\newcommand{\be}[1]{\mid\!{#1}\!\mid}
\newcommand{\no}{\nonumber}
\newcommand{\etal}{{\em et~al }}
\newcommand{\geff}{g_{\mbox{\it{\scriptsize{eff}}}}}
\newcommand{\da}[1]{{#1}^\dagger}
\newcommand{\cf}{{\it cf.\/}\ }
\newcommand{\ie}{{\it i.e.\/}\ }
\title{Enhancement of magnetic anisotropy barrier in long range 
interacting spin systems}
\author{F.~Borgonovi} 
\affiliation{Dipartimento di Matematica e
Fisica and Interdisciplinary Laboratories for Advanced Materials Physics,
 Universit\`a Cattolica, via Musei 41, 25121 Brescia, Italy}
\affiliation{Istituto Nazionale di Fisica Nucleare, Sezione di Pavia, via Bassi 6, 27100 Pavia, Italy}
\author{G.~L.~Celardo}  
\affiliation{Dipartimento di Matematica e
Fisica and Interdisciplinary Laboratories for Advanced Materials Physics,
 Universit\`a Cattolica, via Musei 41, 25121 Brescia, Italy}
\affiliation{Istituto Nazionale di Fisica Nucleare, Sezione di Pavia, via Bassi 6, 27100 Pavia, Italy}
\begin{abstract}
Magnetic materials are usually characterized by
anisotropy energy barriers which dictate the 
time scale of the magnetization decay and consequently the magnetic
stability of the sample.
Here we consider magnetization decay in a class of spin systems 
in contact with a heat bath,
with long range interaction and on site anisotropy.
We show that anisotropy energy barriers can be determined from ergodicity 
breaking energies of the corresponding isolated system, independently of
 the mechanism of magnetic decay (coherent rotation or nucleation). 
In  particular, we show that, in presence of long range 
interaction the anisotropy energy barrier
grows as the particle volume, $V$, for sufficiently large volume, 
and it can grow, for finite sized systems, as $V^{2-\alpha/d}$,
where $\alpha \leq d $ is the range of
interaction and $d$ is the embedding dimension.
As a consequence there is a relevant enhancement of 
the anisotropy energy barrier compared with the short range case,
where the anisotropy energy barrier  grows at most as the particle 
cross sectional area for
large or elongated particles.
\end{abstract}
\date{\today}
\maketitle

\section{Introduction}
A truly comprehensive understanding of magnetism at the nanoscale 
is still lacking. 
From a theoretical point of view the problem of magnetization
decay in nanosystems is difficult to treat:
nanoscopic systems are 
too big to be solved by brute force calculation and too small
to be tackled by the tools of 
statistical mechanics at the equilibrium.
Indeed,  the problem of magnetization decay is a typical
example of out of equilibrium phenomenon, which is the decay out
of a metastable state. 

On the other hand, nanomagnetism  
has also important consequences 
in the technology of memory  and information processing  devices.
The quest for improving magneto-storage density
calls for the realization of smaller and smaller magnetic units.
Significant improvements in experimental techniques 
allowed investigations
of magnetic properties in nanoparticles and nanowires\cite{barbara}.
In particular, recently,  there has been great interest in
Single Chain Magnets (SCM)~\cite{vindigni1,vindigni2,nowak1,gambardella,carbone}, 
which are possible candidates  for nanoscopic memory units.
Nanoscopic systems can also
show ferromagnetic behavior at low but
finite temperature, even if a ferromagnetic phase transition is theoretically 
forbidden~\cite{mermin}, due to large magnetic decay times.

The modelization of the magnetization decay, through an on--site
anisotropic barrier is quite typical in literature, even if, 
mainly short-range
interactions have  been considered.
Nevertheless, in many realistic situations, 
one needs to go beyond nearest neighbor coupling,
taking into account the long range nature of the interaction  
defined by a two--body spin interaction constant decaying at large
distance with a power law exponent $\alpha $ not
larger than the embedding 
spatial dimension $d$\cite{libroruffo}.
It is the case, for instance, of the dipolar interaction
in 3--d systems,
or of the so--called 
RKKY (Ruderman-Kittel-Kasuya-Yosida) interaction,
which decays as  $r^{-d}$, where 
$r$ is  the distance between spins, and $d$ is
the dimension of the lattice system.
In particular, the latter
might be  responsible  for the ferromagnetic behavior of 
Diluted Magnetic Semiconductors (DMS) \cite{dms} and 
Diluted Magnetic Oxides (DMO) \cite{coey}, promising
materials for the realization of spintronics devices.

One of the first attempts to understand magnetic
decay times in nanoparticles is due to Ne\'el\cite{neel}
 and Brown \cite{brown}, who considered that all the spins in
a magnetic particle move coherently, so that they can be considered
as a single spin and described   magnetization decay as  
due to thermal activation over a single energy barrier.
In Brown's theory this time, $\tau$,
is shown to follow an Arrhenius Law : 
\beq
\tau \propto e^{\beta \Delta E}
\label{al}
\eeq
where
$\beta = 1/ k_B T$ is the 
inverse temperature and $\Delta E \propto V $ is the anisotropic energy
barrier  proportional to the particle volume $V$ . 
A step forward Brown's theory has been realized by Braun \cite{braun}.
In his theoretical approach 
a sufficiently elongated  system of short range interacting spins
with an on--site anisotropy barrier,
have been shown to reverse their magnetic moments (thus
producing an average magnetization decay)  through
a process called nucleation, energetically convenient
with respect to coherent rotation.
In this mechanism, accomplished by  the  formation
of a soliton-antisoliton domain wall,
the magnetic anisotropic energy to be overcome 
turns out to be proportional
to the {\it cross sectional area} of the particle,
$\Delta E \propto A$  and not to its  volume $V$. 
Studies of  different mechanisms  of magnetic
decay have been the objective of intense investigation\cite{nowak2} 
until recently, where also 3--d spherical samples with short range interactions
and on--site anisotropy  are shown 
to produce nucleation for sufficiently large radius\cite{nowak3}.
Thus, for short range interaction, Brown's theory, and a consequent 
Arrhenius Law with an exponent proportional to the volume $V$
of the particle is valid only for very small
particles, while in general, for large or elongated particles, 
the exponent is given by the cross sectional area of the particle.
A smaller  exponent means smaller decay
times for the same temperature.
The size and shape dependence of the magnetic anisotropy barrier,
and consequently of the decay times, have been also 
confirmed  experimentally in Ref.~\cite{bode}.

The consequences of long range interaction on magnetic decay 
is much less investigated.  
Long range interaction can affect the decay out of a metastable
state in a significant way; in the seminal paper \cite{griffiths}
it was shown that the decay time out of a metastable state 
in a toy model with infinite range interaction
is given by the Arrhenius Law with an 
exponent proportional to the squared volume  of the particle. 

The main goal of this paper is to analyze  magnetic decay 
beyond nearest-neighbor interaction, focusing on realistic
long range interacting systems. 
In order to estimate the dependence of magnetic decay times on 
temperature, we propose a different point of view,
which turns out to be independent of the  decay mechanism and related
to the recently found Topological Non-connectivity Threshold (TNT)
in anisotropic spin systems~\cite{jsp,tntruffo}.


\section{Topological Non-Connectivity Threshold and Magnetization decay time}
In this Section, following Ref.~\cite{jsp}, we briefly review the Topological Non-Connectivity Threshold.

Let us consider 
a generic  anisotropic spin system,  with an easy axis of magnetization,
(the direction $ \hat{n}_{easy}$ of the magnetization in the ground 
state), with a microcanonical 
energy $$H( \vec{S}_1,\ldots,\vec{S}_N)=E.$$
Let us also set 
\beq
 m = (1/N) \sum_k \vec{S}_k \cdot \hat{n}_{easy}, 
\label{m}
\eeq
as the magnetization along the easy axis. Note that in our paper
it will be $\hat{n}_{easy}=\hat{z} $.

It was proven~\cite{jsp} that below a suitable  threshold, $E_{tnt}$, given by the minimal
energy attainable under the constraint
 of zero magnetization $m$ along the easy axis:
\beq
E_{tnt}= {\rm Min } (H (...\vec{S}_i... )\ 
| \  m =0),
\label{Etnt}
\eeq
 the constant energy surface is disconnected in two portions,
characterized by a different sign of the magnetization.
From the dynamical point of view one has a case of
 ergodicity breaking: a trajectory
at fixed energy cannot change the sign of
magnetization since it is confined forever in one region of the 
phase space.

It was also demonstrated that in case of long range
interaction among the spins\cite {musesti}, 
the disconnected energy portion
determined by the TNT, remains finite, 
when the number of particles becomes infinite. 

While  for isolated systems, the magnetization cannot reverse its
sign if its microcanonical energy $E$ is below the energy
threshold $E_{tnt}$, when the system
is put in contact with a heat bath this may happen 
for any (even extremely low) temperature. Question arise whether
the TNT dictates in some way the magnetization decay.
This is exactly what we have found (see Sections below), namely   
$\Delta E_{tnt}=E_{tnt}-E_{min}$,  represents an effective energy barrier 
for magnetic decay and the decay time depends 
exponentially on such  energy barrier:
\beq
\tau = \tau_0  \ e^{\beta (E_{tnt}-E_{min}) } = 
\tau_0 \ e^{\beta \Delta E_{tnt} },
\label{tau}
\eeq
where, $\tau_0$ is a  factor, that may depend  
on temperature too, $E_{min}$ is the ground state
energy and $\beta=1/k_BT$. While this was confirmed in 
simple toy models with all-to-all interaction among the spins 
\cite{spadafora}, here our aim is    to generalize
these results to realistic spin systems.

It is possible to give a heuristic justification of Eq.~(\ref{tau}).
Magnetic decay occurs through fluctuations of the magnetization
around its equilibrium value. 
The probability a magnetization fluctuation
is determined by the 
free energy barrier, $\Delta F = \Delta E - k_B T \Delta S,$
through the Arrhenius factor, $e^{-\beta\Delta F}$.
Since the entropic barrier,  $\Delta S$, is usually negligible 
at low temperature,
the accessible spin configurations can be determined minimizing the
energy only.    
In order to reverse its sign, the value of the magnetization has to
go, say, from a magnetization $m=1$ to  $m=0$. 
Since for $m=1$ the system is in its minimal
energy, it is clear that $\Delta E_{tnt}$ represents  the minimal
energy barrier found by  the system  while reversing its magnetization. 

\section{The Model : Isotropic $\alpha-$Ranged plus  On--Site Interaction}
Let us now focus on  spin systems with isotropic 
long--range exchange interaction and on--site anisotropy,
described by the following Hamiltonian:
\beq
H = -J \sum_{i>j} 
\frac{ \vec {S}_i \vec {S }_j}{r_{i,j}^{\alpha}} - D \sum_i (S^z_i)^2, 
\label{isoD}
\eeq
where, $ \vec {S_i}$ are the spin vectors with unit length,
$\alpha$ determines the range of the interaction 
among the spins, $J>0$ is the exchange coupling 
and $D>0$ is the on--site energy anisotropy. 
The minimal energy for this class of spin systems is attained
when all the spins are alligned along the $\hat{z}$ direction,
which thus defines the easy axis of the magnetization.

In the following we will mainly focus on the ``critical'' case $\alpha=d$,
since it is relevant for realistic applications (dipole interaction
 in $3$-D and RKKY interactions); 
here critical is referred 
to the fact that interaction is defined as long--ranged for $\alpha<d$,
while it is short ranged for $\alpha>d$~\cite{libroruffo}.
Moreover, we choose Hamiltonian~(\ref{isoD}) 
for sake of  generality since we would like to consider 
not a single specific interaction, but a class of interactions 
depending as some power of distance among spins.

For this class of systems, 
the energy $E_{tnt}$, see Eq.~(\ref{Etnt}), can  be computed numerically 
using a minimizing constrained algorithm.
We can also estimate analytically  $E_{tnt}$
for a generic range of the interaction $\alpha$.
To this purpose let us consider two configurations with
$m=0$: 
\begin{itemize}
\item the first one with all 
spins  aligned perpendicular to the easy axis.
 The energy difference of this configuration 
from the one having minimal energy is  $DN$,
which is the energy barrier due to the coherent rotation of all
spins; 
\item  
a configuration, labelled $\uparrow\downarrow$,
consisting ot two neighbors identical blocks with opposite magnetization
along the easy axis.
This configuration roughly corresponds
to what is called nucleation configuration in literature.
The energy difference of this configuration from the minimal energy
in the case $\alpha \leq d$,  is given by~\cite{musesti}:
\begin{equation}
\label{tnt}
\Delta E_{\uparrow\downarrow} =  J C_{\alpha, d} N^{2-\alpha/d}, 
\end{equation}
where $C_{\alpha, d}$ is a suitable constant, for instance, 
$C_{1,1} \simeq 4 \ln 2 $.
\end{itemize}

\begin{figure}[!ht]
\vspace{0cm}
\includegraphics[width=8cm]{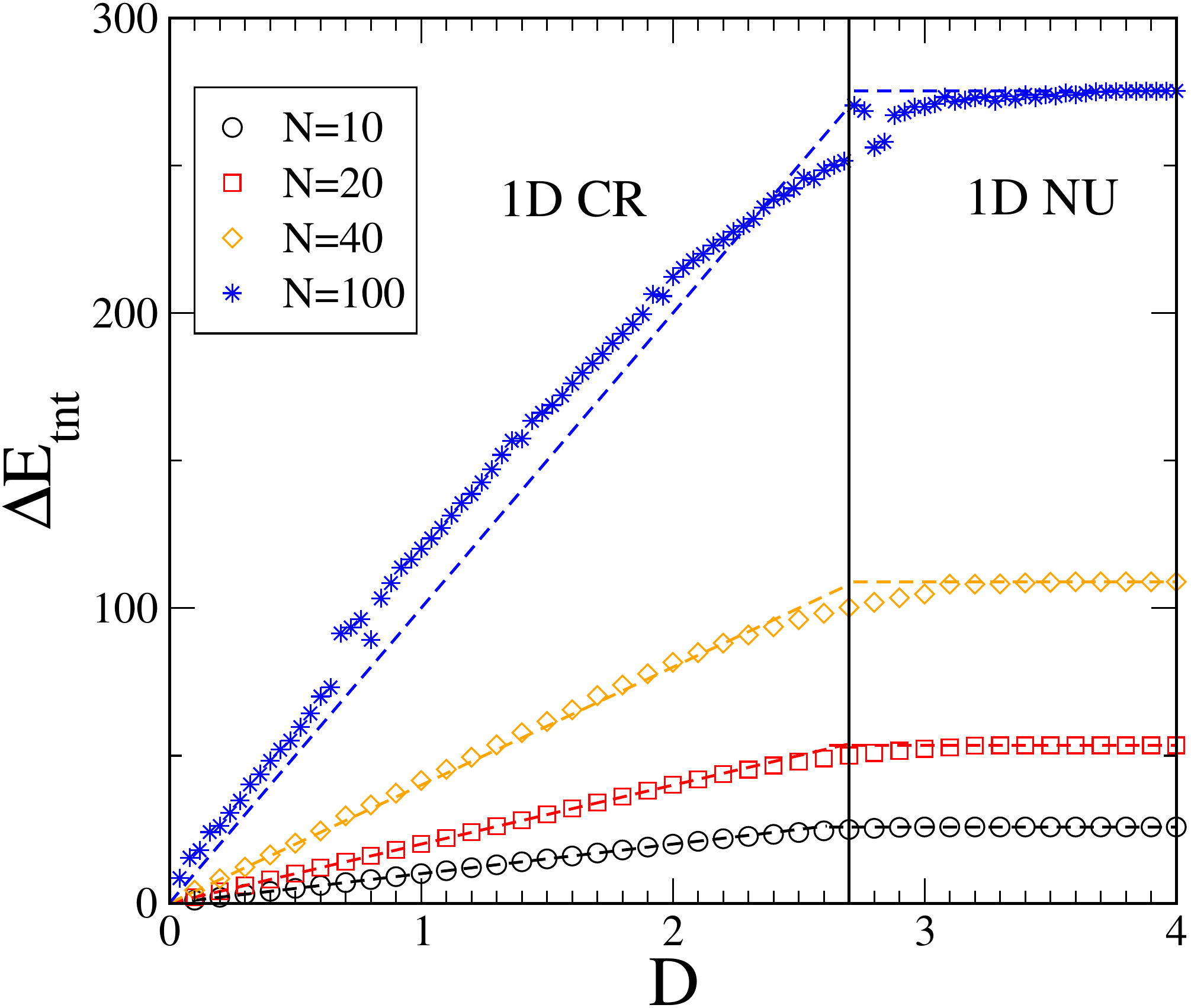}
\caption{Energy barrier as a function of the anisotropy
barrier $D$ for the $\alpha=d=1$.
Different symbols indicate different number of spins in the linear array, 
as indicated in the legend. 
Dashed lines indicate the approximation Eq.~(\ref{ddee}).
The vertical black line indicates $D=D_{cr}$, and 
separates the region of coherent rotation  (CR, left), 
from the region of nucleation (NU, right).
}
\label{tntfig}
\end{figure}

Following similar reasoning as in Ref.~\cite{musesti}, 
it can be shown that the energy of these two configurations is a good
approximation of $E_{tnt}$, so that we can write
\begin{equation}
\Delta E_{tnt} \equiv E_{tnt} -E_{min} \approx {\rm Min} (DN, \Delta E_{\uparrow\downarrow}),
\label{ddee}
\end{equation}
Note that Eq.~(\ref{ddee}), valid whenever $DN$ is not close to
$\Delta E_{\uparrow \downarrow}$,
gives an estimation of the anisotropic energy barrier, which
can be used in Eq.~(\ref{tau}) to get magnetic decay times.
Eq.~(\ref{ddee}) has been confirmed numerically in Fig.~(\ref{tntfig}), 
for the case $\alpha=d=1$, and we have found coherent rotation 
for $D \ll D_{cr}= 4J \ln 2 $  
 and nucleation for $D \gg D_{cr}$, where the $D_{cr}$ has been obtained  
comparing  Eqs.~(\ref{tnt}) and (\ref{ddee})~\footnote{
Even if it not obvious, a priori, that for long range
interactions, the mechanism of magnetization decay can be
described by the process of nucleation and coherent rotation, we 
assume them to be true and we verify, in the last Section,
that they really occurs in our case.}.
In this figure we compute numerically the energy barrier $\Delta E_{tnt}$
using a minimizing constrained algorithm (symbols) and we compare it with 
the approximation
(\ref{ddee}) for one dimensional chains with 
different number of spins. 
As one can see, the agreement is fairly good.
Note that in case of nucleation 
($DN > \Delta E_{\uparrow\downarrow}$),
the effective energy barrier $\Delta E_{tnt}$
to be  put in the Arrhenius Law
should be increased by a term $D$, called single-spin
flip. This extra term should be added only for
continuous models,
as the modified Heisenberg model we are considering, 
so that, $\tau_0 \propto e^{D \beta}$ \cite{vindigni2}.
This should be done since a single 
spin can continuously change from $S=1$ to $S=-1$
only passing the state   $S=0$. 
While the two extreme states have the same magnetic
anisotropy (since it is proportional to $DS^2$) the intermediate state $S=0$ 
has a  barrier higher by a factor $D$.
This additional spin-flip does not occur, for instance,
 in the Ising--like models,
since there are no intermediate states between   $S=1$  and $S=-1$.

In the following
we analyzed the magnetic decay time in the canonical ensemble, using
a modified Monte Carlo  
simulation \cite{metropolis,nowak2}.
As initial condition we choose all spins aligned along the easy axis,
and from the exponential decay of the average magnetization,
$\langle m(t) \rangle \propto e^{- t/ \tau}$, we
computed the magnetic decay time, $\tau$.

\section{Enhanced Barrier}
In this Section, we show our numerical results for the magnetization
decay. They will clearly give evidence of the enhancement of
the magnetization decay time due to long range interaction.

\subsection{ One dimensional case }
Here we focus on the  one dimensional critical case. 
\begin{figure}[!ht]
\vspace{0cm}
\includegraphics[width=7cm]{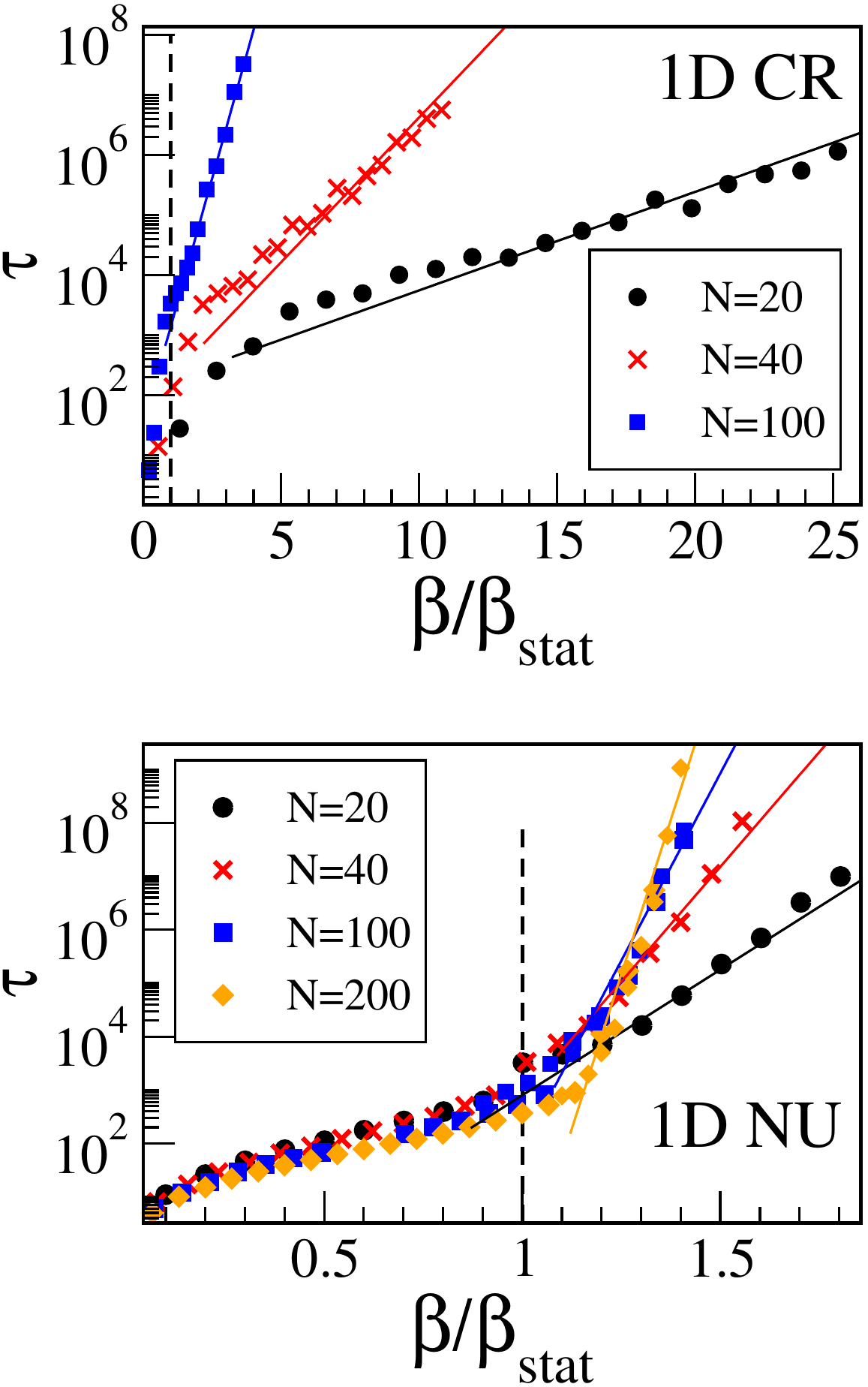}
\caption{Magnetic decay times, $\tau$  {\it vs} the rescaled inverse
temperature, $\beta/\beta_{stat} $ for $\alpha=d=1$ and different numbers
of spins $N$ as indicated in the legend.
Upper panel is  characterized by
coherent rotation  with parameters  $J=1$ and $D=0.05$ , while the lower by 
nucleation with    $J=1/16$ and $D=0.5$.
Different  symbols refer
to numerical results,
while 
full  lines in   are the theoretical predictions
$\exp (\beta \Delta E_{tnt} )$. Specifically for coherent
rotation,   we have $\Delta E_{tnt}=DN$, while
for nucleation,  we have  
$\Delta E_{tnt}= \Delta E_{\uparrow\downarrow} +D$
(see text for explanations).
Vertical dashed lines  refer to the inverse rescaled 
statistical temperature $\beta=\beta_{stat}$ obtained by
a standard mean field approach.
}
\label{uno}
\end{figure}
Our results for the magnetic decay time, are collected in Fig.~\ref{uno},
where we consider the case $\alpha=d=1$:  magnetic decay times 
are shown in the upper panel 
for coherent rotation  ($D \ll D_{cr}$)  
 and in the lower panel for nucleation ($D \gg D_{cr}$).  
Samples with different
number of particles experience different anisotropy energy barriers. As one
can see the numerical results indicated by symbols well agree with the
theoretical prediction given by Eq.~(\ref{ddee}), shown as full lines.
Our results for coherent rotation, Fig.~\ref{uno} (upper panel),
clearly indicates 
that Brown's theory of coherent rotation with an anisotropy energy barrier 
proportional to the particle volume ($V \propto N$ ) is still at work 
even for long range interacting systems.
In Fig.~\ref{uno} (lower panel), 
the case of nucleation is shown. Also in this case
the  anisotropy energy barrier is proportional to the particle volume.
Let us remark that this behaviour is different 
for short range interacting systems,
where in the one dimensional case, the anisotropic energy barrier is
independent from the size of the system~\cite{vindigni2,carbone}.
Note that for high temperature (small  $\beta $ values)
symbols with different number of spins lye upon the same curve, that turns
out to be almost independent of $N$.

\subsection{Multidimensional case}

Our results can be  extended to higher dimensions. 
In particular we will show that for large enough particle,
the anisotropic energy barrier is always proportional
to the volume of the particle. Moreover we point out
that for finite sized systems, the anisotropic energy barrier can
grow even faster than the volume. 

From  Eqs~(\ref{tnt})  and~(\ref{ddee}), we have that 
while for coherent rotation, the anisotropy energy is always proportional
to the volume of the particle, for nucleation the anisotropy energy
can grow even faster than the volume of the particle,
$\Delta E_{\uparrow\downarrow}  \propto N^{2-\alpha/d}$.
Since the anisotropy energy barrier is determined by the smallest of the two barriers, it follows that
for sufficiently large particles  the energy
barrier will always be proportional to the particle volume
since $V \propto N$. 
In particular, this means that even in the case of 3--d elongated particles,
differently with what happens in the short range case, 
the energy barrier {\it does not
depend on the cross sectional area but on the whole
particle  volume.}

\begin{figure}[!ht]
\vspace{0cm}
\includegraphics[width=7cm]{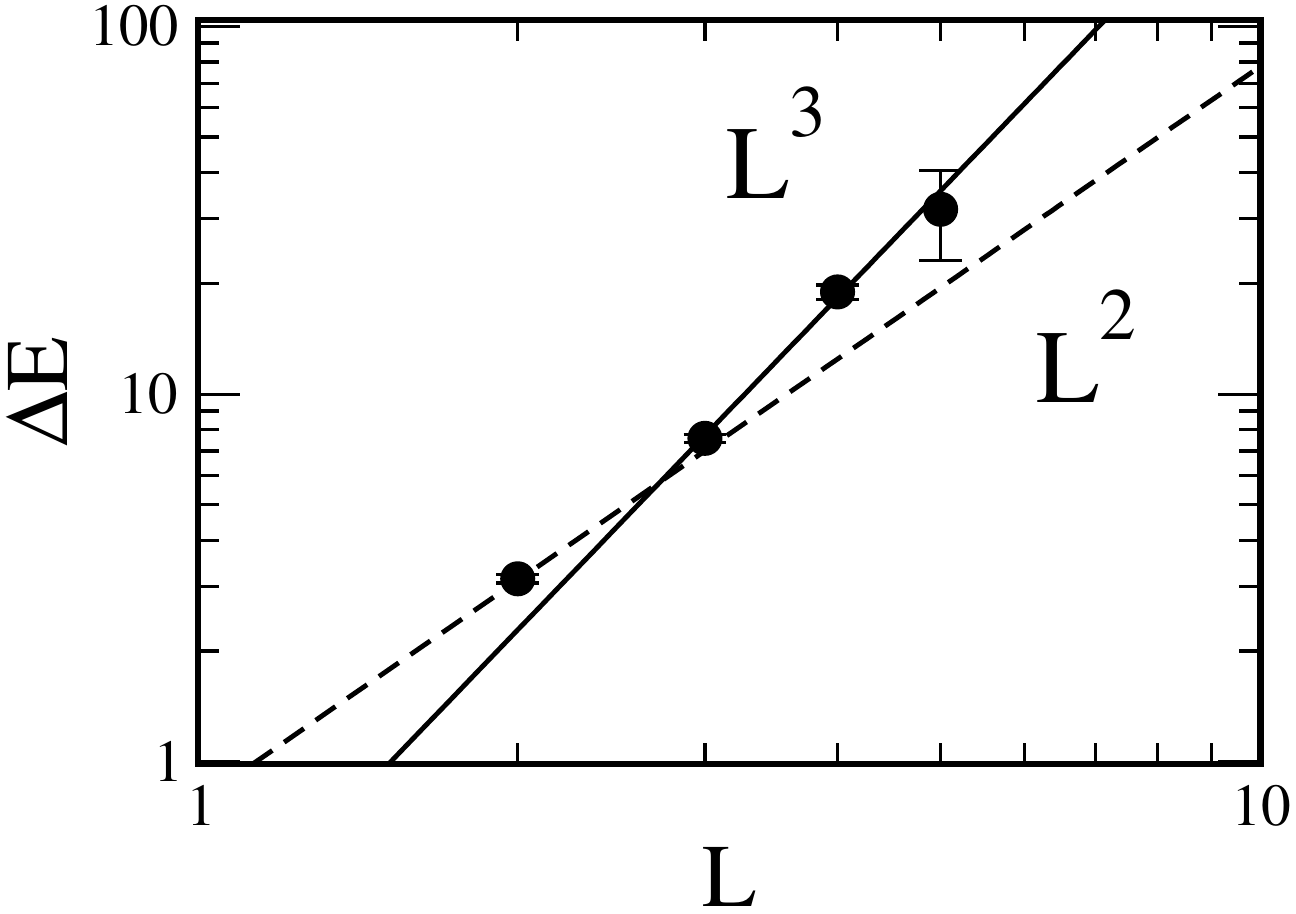}
\caption{Energy barrier $\Delta E$ as a function
of the side  $L$ of the parallelepiped, obtained
by an exponential fitting of the decay time {\it vs} the inverse temperature $\beta$.
Here is 
$\alpha=3$,
$J=1/20$ and $D=0.5$ (nucleation).
}
\label{fin}
\end{figure}

This theoretical prediction has been numerically confirmed in Fig.~\ref{fin},
where we study the case of nucleation in thin parallelepipeds
$L \times L \times \epsilon L $,  in presence of a `` critical''  interaction $\alpha=d=3$.
 We chose the aspect ratio $\epsilon = 10 $. 
 
The energy barrier $\Delta E$ extracted from the numerical fit of the
Arrhenius Law $\tau = \tau_0 \exp(\beta \Delta E)$
has been plot as a function of the side $L$
of the parallelepiped. For sake of comparison
the lines proportional to the cross section
($L^2$) and to the volume ($L^3$) has been
drawn. As one can see without any doubt the barrier grows as the volume.

\begin{figure}[!ht]
\vspace{0cm}
\includegraphics[width=7cm]{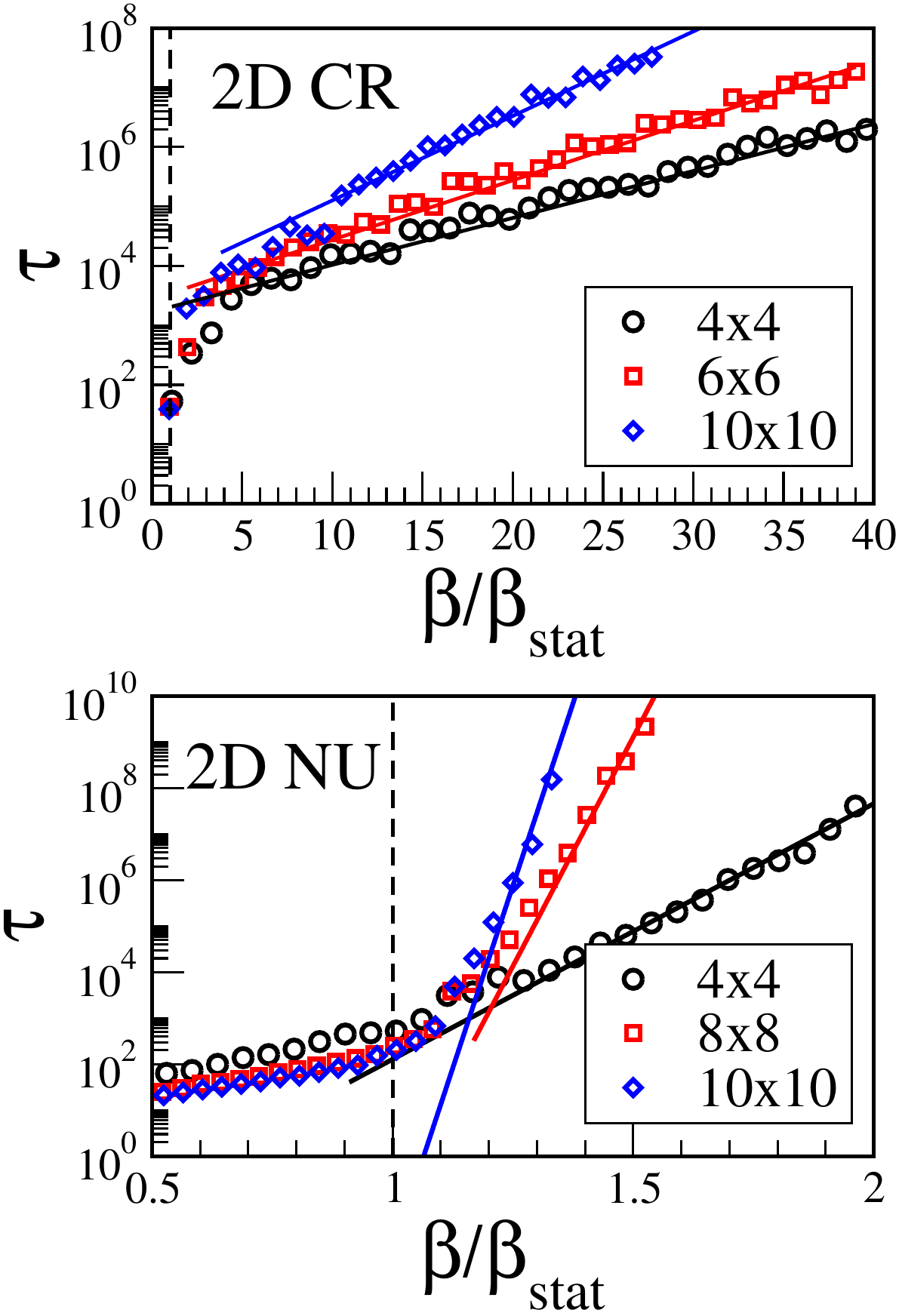}
\caption{Decay time $\tau$
{\it vs} the rescaled inverse temperature $\beta/\beta_{stat}$  for a
square lattice  ($d=2$) with $\alpha=1$.
for  $J=1$ and $D=0.025$ (upper panel, coherent rotation);
and  for $J=1/40$ and
$D=1/2$ (lower  panel, nucleation).
Symbols refer to numerical data, while full lines are the
analytical prediction  Eq.~(\ref{ddee}).
Vertical dashed lines represent the rescaled inverse
statistical temperature  $\beta/\beta_{stat}=1$.
Note that in case of coherent rotation (upper panel),
the anisotropy energy is proportional to $DN$, while in the case of nucleation
(lower panel), the anisotropy energy is proportional to $JN^{3/2}$.
Lattice dimensions have been indicated in the legend}
\label{fisoD}
\end{figure}
From  Eqs~(\ref{tnt})  and~(\ref{ddee}),  it also follows
 that, when $\alpha < d$ the
anisotropy barrier can grow even faster than the volume of the particle.
This can happen up to a critical value of the particle volume, for which $DN=\Delta E_{\uparrow \downarrow}$.
Such an effect is a distinguished feature 
of long--range interaction
and it is  forbidden in the case of short range interactions.
To confirme such theoretical prediction let us consider a square lattice
($d=2$) characterized by an exchange interaction decaying as $1/r$ ($\alpha=1$).
Even in this case we can distinguish between the regime of coherent rotation
(large anisotropy barriers $D$) from that of nucleation 
(small $D$ values).
The results of our simulation have been shown 
in  Fig.~\ref{fisoD}. 
In the lower panel the theoretical predictions have been indicated 
as full lines while symbols represents numerical results from
Montecarlo simulation. As one can see the agreement with the theory,
which predicts an energy barrier $\Delta E_{tnt} \propto J N^{3/2}$
is fairly good. For sake of comparison, the regime of coherent
rotation (Fig.~\ref{fisoD}, upper panel) is also shown, indicating that,
in this case,
 the energy barrier  is proportional to the volume of the particle, namely  $\Delta E_{tnt} \propto D N$.

Another important point  to be discussed is in which temperature range the Arrhenius Law 
with the exponent given by Eq.~(\ref{ddee}) holds. 
In the cases of coherent rotation and nucleation, 
see also Ref.~{\cite{brown}}, we might expect that the Arrhenius Law 
is valid only when   $k_B T \ll  \Delta E_{tnt}$.
Clearly this gives an upper bound for the 
temperature  for which Eq.~(\ref{tau})
is valid. Moreover it should be $T \ll T_{stat}$,
where the latter is the temperature at which the barrier at $m=0$
in the free energy vanishes (and it coincides with the 
temperature at which a phase transition occurs in the thermodynamic
limit). It is very interesting that, computing 
$T_{stat}= 1/k_B  \beta_{stat}$ by means of
a standard mean field approach, one gets a very nice estimate 
of the validity range of the Arrhenius Law (see dashed vertical lines in Figs.
\ref{uno} and \ref{fisoD}). Nevertheless we stress that this delicate point 
needs further analysis to be clarified.

\section{Coherent Rotation and Nucleation}
\label{sec:Pm}
\begin{figure}[!ht]
\vspace{0cm}
\includegraphics[scale=0.5]{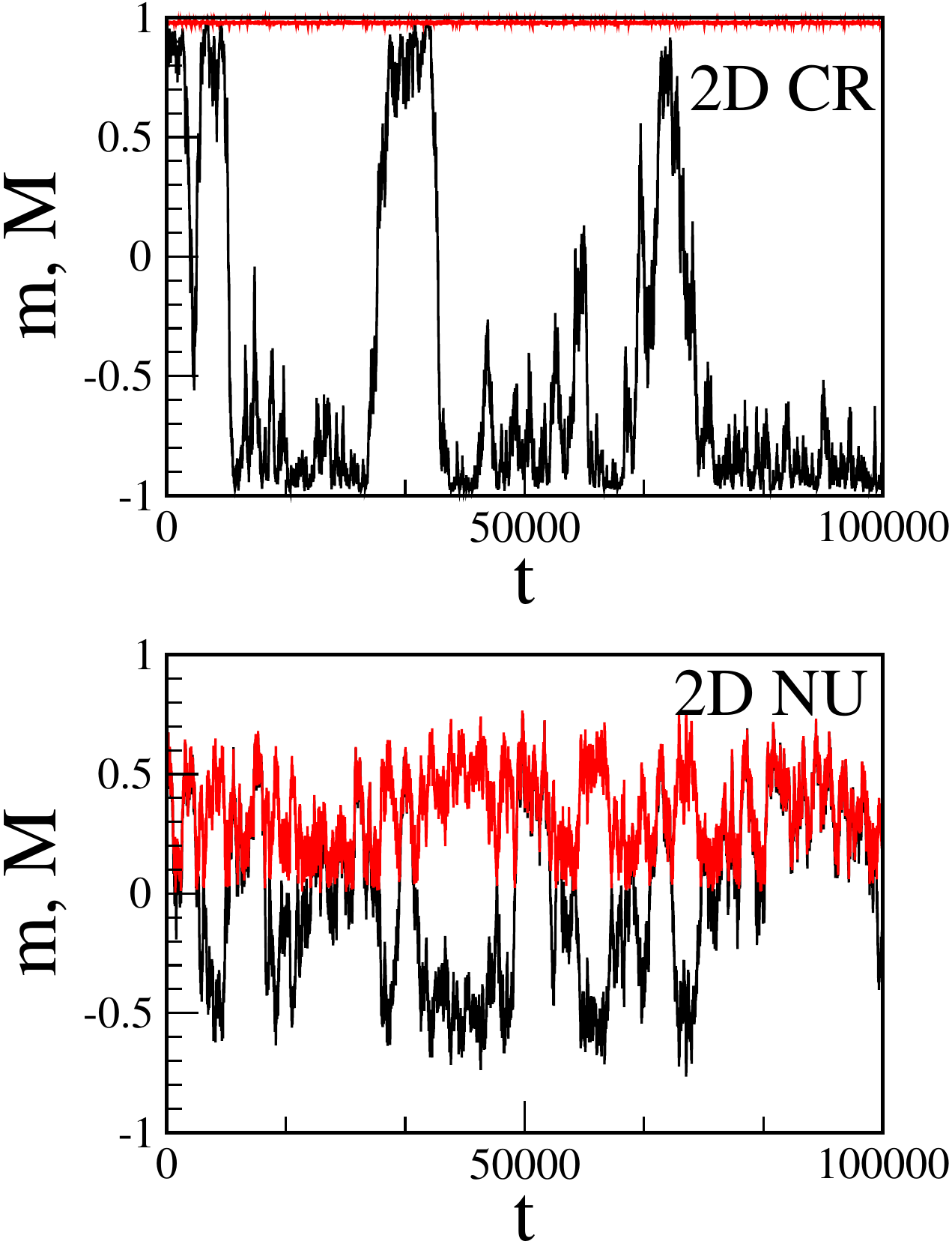}
\caption{Upper panel (coherent rotation) for a square lattice 
$6 \times 6$, $\beta=5$. 
Lower panel (nucleation) for a square lattice $8 \times 8$
 and $\beta=4.5$. In both panels 
the red  curve refers to $M$, while the black one  to $m$.
}
\label{last}
\end{figure}

Let us notice that the use of the words coherent rotation
and nucleation, are not only ways to label two different mechanisms taken
from the short range models, but  correspond to the spin dynamics.
To prove that we first consider in Fig.~\ref{last} two different situations 
taken from the upper (coherent rotation) and the lower panel (nucleation)
of Fig.~\ref{fisoD}. In particular we consider one single Montecarlo trajectory
and we plot, in Fig.~\ref{last}  the value of $M = |\frac{1}{N} \sum_{i=1}^N \vec{S}_i|$ and $m=\frac{1}{N} \sum_{i=1}^N S^z_i$,
as a function of the Metropolis time. It is clear that while in the case
of coherent rotation (upper panel of Fig.~\ref{last})
$M \approx 1 $ and $m$ change its sign,  
 for nucleation (lower panel of Fig.~\ref{last})
one has  $M \simeq |m|$.

As an interesting consequences of the coherent motion of the spins
 during the magnetic reversal, we show here,
focusing   on the one dimensional case, that one can 
 estimate the whole probability distribution of the 
magnetization in the canonical ensemble, $P_T(m)$ at low temperature 
and around $m=0$. 

Note that the  knowledge of the probability distribution of 
the magnetization allows one to compute
the magnetic decay time: 
\begin{equation}
\tau \propto e^{\beta\Delta F} = \frac{P_T(m_{max})}{P_T(m_{min})},
\label{ddf}
\end{equation} 
where
$P_T(m_{max,min})$ stands for the maximum/minimum 
value of the probability distribution. 

Due to  the coherent motion of all spins it is
possible to express the energy of the system as a function 
of the total magnetization along the easy axis; from the 
knowledge of $E(m)$  we can compute   $P_T(m)$ since:
$$
P_T(m) \ dm= P_T(E) \ dE = e^{-E(m)/k_B T} \rho(E) \ dE 
$$
where $\rho(E)$ is the density of states at fixed energy. 
Using the fact that
$\rho(E) dE= \rho(m) dm$, where  $\rho(m)$ is the density of
 states at fixed magnetization,
we have: 
\beq
P_T(m) \  dm = e^{-E(m)/k_B T}\  \rho(m) \ dm. 
\label{pTm}
\eeq
Let us  compute $E(m)$ for coherent rotation and nucleation.
When the decay of the magnetization occurs through
coherent rotation we have $\vec{S}_i \vec{S}_j=1$, and 
$S_i^z= \langle S^z \rangle=m$
so that, from Eq.~(\ref{isoD}) we have:
\beq
E (m) = -J/2 \sum 1/r_{i,j}^{\alpha} -DNm^2.
\label{Ecr}
\eeq

Since all spins move coherently on a sphere,
the density of states at fixed magnetization along any axis is constant, 
$\rho(m)= const.$ Moreover 
the first term in the r.h.s. of Eq.~(\ref{Ecr})
 is independent of $m$  and we have (for coherent rotations):  
\beq
P_T(m) \propto e^{\beta DNm^2} .
\label{Pcr}
\eeq

\begin{figure}[!ht]
\vspace{0cm}
\includegraphics[width=7cm]{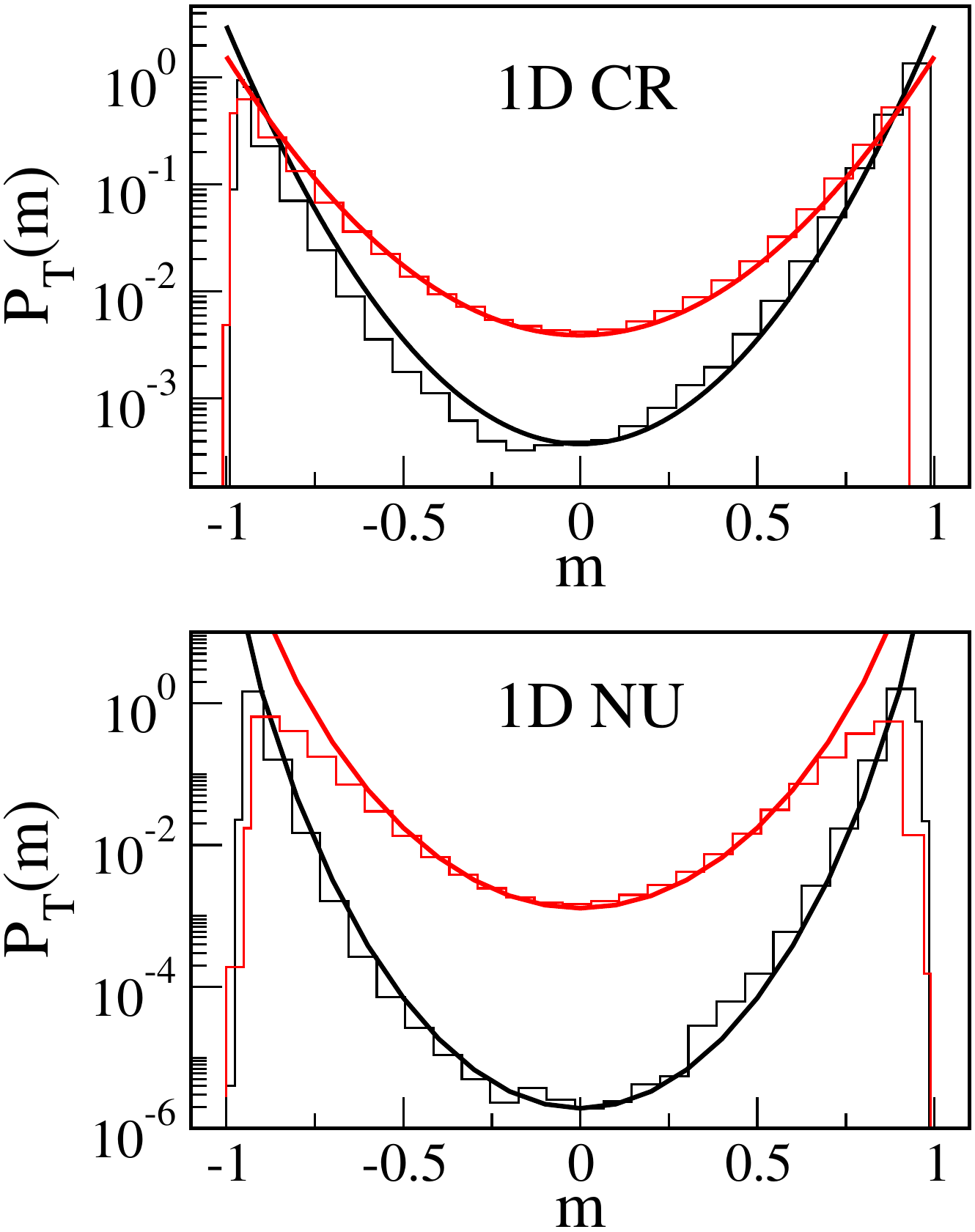}
\caption{
Probability distribution of the magnetization $P_T(m)$ {\it vs}
the magnetization $m$ for $\alpha=d=1$, $N=20$ and
different temperatures $\beta=1/k_{B}T$.
Upper panel is for   $J=1$ and $D=0.05$  (coherent rotation regime)
and
$\beta=6$ (upper red hystogram), $\beta=9$ (lower black hystogram).
The lower panel is 
for $J=1/16$ and $D=0.5$ (nucleation)
and 
$\beta=8$ (upper red hystogram), $\beta=11$
(lower black hystogram).
Hystograms represent
numerical data, while smooth curves analytical results.
}
\label{unob}
\end{figure}

We verified numerically that Eq.~(\ref{Pcr}) gives
an excellent approximation of the probability distribution of the 
magnetization around $m=0$ in the 
coherent rotation regime, see Fig.~\ref{unob} (upper panel).

In case of nucleation  the possible configurations which the system can visit
can be obtained assuming that magnetic decay occurs
by first reversing one of the edge spins, then its nearest neighbor,
and then the spin immediately after until
 all spins are reversed. 
If we have $k$ spins on the left pointing upwards and $N-k$ spins
pointing downwards, we can write the energy of this configuration
as 
\begin{equation}
\label{ebb}
E(k,N)= E_{min}(k) + E_{min}(N-k)+W(k,N-k),
\end{equation}
 where  $W(k,N-k)$ is the 
interaction energy between the two neighbors 
blocks of $k$ and $N-k$ spins and
\begin{equation}
\label{ebb1}
E_{min}(k)= -J/2 \sum^k_{i=1} \sum^k_{j \ne i} 1/r_{i,j}^{\alpha} -Dk,
\end{equation} 
is the minimal energy for a system of $k$ spins.
The minimal energy of a system of $N$ spins can be easily obtained 
from (\ref{ebb}) by changing the sign of $W(k,N-k)$,
\begin{equation}
E_{min}(N)=E_{min}(k) + E_{min}(N-k)-W(k,N-k).
\end{equation}
Taking into account that $m=(2k-N)/N$ is the magnetization of 
a system composed
by $k$ spins along the easy axis and $N-k$ opposite to it, 
one gets,
\begin{equation}
E(m) = 2E_{min}(N_+)  +2E_{min}(N_-) - E_{min}(N),
\label{newen}
\end{equation}
where $N_{\pm}= N(1-m)/2$.
In this case  it is more difficult
to estimate $\rho(m)$. Assuming, also in this case
 $\rho(m)=const.$, from 
the knowledge of $E(m)$ (Eq.~(\ref{newen})), 
we  obtain  $P_T(m) \propto e^{-\beta E(m)} $
which  agrees very well (apart close to $m= \pm 1$) 
with numerical results in the  nucleation regime,
see Fig.~\ref{unob} (lower panel). 

The validity of Eq.~(\ref{tau}) and of our method
to approximate $P_T(m)$ at low temperature, have been
tested for a wide range of values of $\alpha$
(even for short range interactions) and
a detailed study will be reported elsewhere \cite{fausto}.

\section{Conclusions}
In conclusion we propose a general method to determine 
the anisotropic energy barrier in spin systems. 
The barrier, which determines magnetic decay times, can be computed 
from the topological non-connectivity energy threshold in the corresponding
isolated systems. Our analysis shows  that  for long
range interaction two different regimes can be identifyied, coherent
rotation and nucleation.
In the first regime we have predicted and numerically confirmed
that magnetic decay time depends exponentially on the volume of the 
particle. 
Nevertheless, in the regime of nucleation, the magnetic barrier
depends on a suitable power of the volume, $V^{2-\alpha/d}$,
where $\alpha < d $ is the range of interaction and $d$
is  the lattice dimension.
These facts have two remarkable consequences. The first one is that
magnetic decay times  depend exponentially on the volume of the 
particle and not on its cross sectional area, as happens
for short range interactions.
The second one is that, the fast growth of the magnetic barrier
with the volume, gives  rise to the possibility
to observe long-lived metastable ferromagnetism in finite systems
at high temperature which
is ruled out for short range interaction.

\hspace{1.cm}

We acknowledge useful discussion with B. Goncalves, D.~Mukamel,
S.~Ruffo, R.~Trasarti-Battistoni and A.~Vindigni.
This work has been supported by Regione Lombardia and CILEA Consortium 
through a Laboratory for Interdisciplinary Advanced Simulation (LISA) 
Initiative   (2010)
 grant [link:http://lisa.cilea.it] ". Support by the grant D.2.2 2010
(Calcolo ad alte prestazioni)  from Universit\'a  Cattolica 
 is also acknowledged.

\end{document}